# Design of magnetic spirals in layered perovskites: extending the stability range far beyond room temperature

Tian Shang[1,2,3,*], Emmanuel Canévet[4,5], Mickaël Morin[1,6], Denis Sheptyakov[4], María Teresa Fernández-Díaz[7], Ekaterina Pomjakushina[1], and Marisa Medarde[1,*]

[1] Laboratory for Multiscale Materials Experiments,

Paul Scherrer Institut, CH-5232 Viligen PSI, Switzerland

[2] Swiss Light Source,

Paul Scherrer Institut, CH-5232 Viligen PSI, Switzerland

[3] Institute of Condensed Matter Physics,

École Polytechnique Fédérale de Lausanne (EPFL), CH-1015 Lausanne, Switzerland

[4] Laboratory for Neutron Scattering and Imaging,

Paul Scherrer Institut, CH-5232 Viligen PSI, Switzerland

[5] Department of Physics, Technical University of Denmark, 2800 Kgs. Lyngby, Denmark

[6] Excelsus Structural Solutions (Swiss) AG

PARK innovAARE, CH-5234, Villigen, Switzerland

[7] Institut Laue Langevin,

71 avenue des Martyrs, CS 20156 -38042 Grenoble CEDEX 9, France

*Corresponding authors (tian.shang@psi.ch; marisa.medarde@psi.ch)

## Abstract

In insulating materials with ordered magnetic spiral phases, ferroelectricity can emerge due to the breaking of inversion symmetry. This property is of both fundamental and practical interest, in particular with a view to exploiting it in low-power electronic devices. Advances towards technological applications have been hindered, however, by the relatively low ordering temperatures $T_{spiral}$ of most magnetic spiral phases, which rarely exceed 100 K. We have recently established that the ordering temperature of a magnetic spiral can be increased up to 310 K by the introduction of chemical disorder. Here we explore the design space opened up by this novel mechanism by combining it with a targeted lattice control of some magnetic interactions. In Cu-Fe layered perovskites we obtain $T_{spiral}$ values close to 400 K, comfortably far from room temperature and almost 100 K higher than using chemical disorder alone. Moreover, we reveal a linear, universal relationship between the spiral's wave vector and the onset temperature of the spiral phase. This linear law ends at a paramagnetic-collinear-spiral triple point, which defines the highest spiral ordering temperature that can be achieved in this class of materials. Based on these findings, we propose a general set of rules for designing magnetic spirals in layered perovskites using external pressure, chemical substitutions and/or epitaxial strain, which should guide future efforts to engineer magnetic spiral phases with ordering temperatures suitable for technological applications.

# Introduction

Magnetic frustration is characterized by the magnetic interactions that cannot be simultaneously satisfied[1, 2, 3, 4, 5, 6] and may lead to non-trivial magnetic orders[7, 8, 9, 10, 11]. An example of such exotic magnetically ordered phases is the spin spiral, which has received a lot of attention during the last decade due to its potential for inducing ferroelectricity in insulating materials[12, 13, 14, 15, 16]. A necessary condition to use this property for magnetoelectric applications is to stabilize the spiral state close to room temperature (RT). Unfortunately, very few insulators display spiral phases above 100 K.

Besides cupric oxide (CuO) and some hexaferrites [17, 18, 19, 20], the layered perovskite $YBaCuFeO_5$ is one of the few exceptions to this rule. However, what makes this last material truly remarkable is the extraordinary tunability of its spiral ordering temperature ($T_{spiral}$), which can be increased by more than 150 K by manipulating the degree Cu/Fe chemical disorder in the structure. We showed recently that $T_{spiral}$ can be shifted from 154 K to 310 K by adjusting the average Cu/Fe occupations $n_{Cu}$ and $n_{Fe}$ of the square-pyramidal sites in the crystal unit cell (Fig. 1A) [21]. We also established the existence of a positive correlation between the spiral ordering temperature and the degree of Cu/Fe intermixing (maximal for $n_{Cu} = n_{Fe} = 50\%$). However, the most surprising observation was that tiny differences in the Cu/Fe occupations ($|n_{Cu} - n_{Fe}| \leq 6\%$) can shift $T_{spiral}$ by more than 150 K. Such a huge, positive impact of the Cu/Fe disorder in the spiral stability is remarkable, and calls for further investigations aimed to both, understanding and exploiting this unusual trend.

The aim of the present study is to explore the design space opened by this novel, disorder-based spiral control mechanism. After having shown its potential and limitations for the particular case of YBaCuFeO$_5$ [21], we combine it here with a targeted lattice tuning of some magnetic exchanges. The idea behind is to add-up the effect of both mechanisms in order to stabilize spin spirals at temperatures higher than those obtained using these two strategies separately. In the following we show that for Cu-Fe based layered perovskites, $T_{spiral}$ values close to 400 K can be obtained using this approach. These values are ~100 K higher than using chemical disorder alone[21], and expand the stability range of the spiral to a comfortable temperature region that extends well beyond RT. We also uncover the existence of an intriguing paramagnetic–collinear-spiral triple point which defines the highest $T_{spiral}$ value that can be achieved in this class of materials. Moreover, we find several correlations between the spiral properties and some structural parameters that can be summarized in the form of a simple set of rules for magnetic spiral design in layered perovskites. Besides overcoming limiting factors in terms of operating temperatures, these results could contribute to increase the number of materials featuring spiral phases stable well beyond RT and be an important step towards the technological application of magnetic spirals in spintronics devices.

## Results

**Magnetic interactions**

The strategy followed to attain our goal is based in the particularities of the YBaCuFeO$_5$ crystal structure and the associated nearest-neighbor (NN) exchange couplings. A schematic representation of the layered perovskite unit cell of this material (space group *P4mm*) is shown in Fig. 1A[22, 23, 24]. The A-sites are occupied by equal amounts $Y^{3+}$ and $Ba^{2+}$ which order in planes perpendicular to the *c*-axis owing to their very different ionic radii[25]. The B sites host $Fe^{3+}$ and $Cu^{2+}$ in square-pyramidal coordination that form slabs of 'bowties' separated by $Y^{3+}$ layers. Contrarily to Y and Ba, Cu and Fe display a very particular kind of correlated disorder which is characterized by the preferential occupation of the bowties by Cu-Fe pairs[24]. Such Cu-Fe 'dimers' are disordered in the structure (see Fig. 1B), leading to variable, preparation-dependent averaged Cu/Fe occupations of the atomic positions inside the pyramids.

The Cu-Fe dimers have very important consequences for the magnetism of $YBaCuFeO_5$. As demonstrated in refs. 24, 26, all possible in-plane NN interactions $J_{ab}$ (Fe-Fe, Cu-Cu and Cu-Fe) are strong and antiferromagnetic (AFM), see Fig. 1B. $J_{c1}$, one of the two NN couplings along *c*-axis (Fig. 1C) is also AFM and disorder-independent, but its value is about one order of magnitude smaller. The intra-bowtie coupling $J_{c2}$, comparable to $J_{c1}$ in absolute value, is the only ferromagnetic (FM) NN exchange interaction, and it is worth noting that this is only possible if the bowties are preferentially occupied by Cu-Fe pairs (i.e., if $J_{c2} = J_{Cu-Fe}$)[24]. This set of magnetic interactions is at the origin of the collinear magnetic phase observed at high temperatures ($T < T_{collinear}$) with $\mathbf{k_c}$ = (½ ½ ½), see Fig. 1C[21, 24, 27, 28, 29, 30]. Note that in spite of the underlying chemical disorder, the preferential occupation of the bowties by FM Cu-Fe dimers and the non-dependence of the sign of $J_{ab}$ and $J_{c1}$ with the Cu/Fe disorder imply that the magnetic structure can be seen as a set of *ordered* layers with strong in-plane AFM bonds coupled though weak, alternating FM and AFM bonds along *c*-axis (Fig. 1D).

At low temperatures the $k_z$ component of the propagation vector becomes incommensurate and the collinear magnetic order transforms in an inclined circular spiral with $\mathbf{k_i}$ = (½ ½ ½±$q$) (Fig. 1E)[21, 24, 28, 29, 31]. Although the *c*-axis seems naturally more prone to a magnetic instability due the weakness of $J_{c1}$ and $J_{c2}$, the appearance of spiral state is surprising because these two couplings are not frustrated, and the next-nearest-neighbor (NNN) interactions, estimated from DFT calculations, are either too small, or have the wrong sign to produce frustration along this direction[24, 26, 32, 33].

**Tuning $T_{spiral}$ with chemical disorder**

Very recently, an alternative, disorder-based mechanism has been proposed by Scaramucci and co-workers [32]. These authors demonstrated that a magnetic spiral can be stabilized through the introduction of few strong, randomly distributed frustrating bonds in systems featuring a *single* crystallographic direction with weak NN magnetic couplings. YBaCuFeO$_5$, with strong in-plane couplings and weak, alternating AFM/FM exchanges along *c* (Fig. 1B), clearly fulfills these premises[24, 26]. The required frustrating bonds are provided by small, preparation-dependent amounts of bowties occupied by Fe-Fe pairs (accompanied by the same amounts of Cu-Cu pairs if we assume a constant Cu/Fe ratio). Although both are energetically unfavorable[24], their presence at very low concentrations cannot be disregarded in real materials. Rather, the monotonic changes in the size of the bowties with increasing cooling rates suggest that their concentration could be directly linked to the degree of Cu/Fe disorder. Due to the small $Cu^{2+}$ moment (S = ½) and the extremely weak Cu-Cu coupling ($J_{Cu-Cu}$ ~ 0) (dashed-bonds in Fig. 1D), a Cu-Cu defect is expected to have only a minor impact in the alignment of the neighboring spins. However, an AFM Fe-Fe pair (S = 5/2, $J_{Fe-Fe}$ ~ -100 meV) (solid-green bonds In Fig. 1D) constitutes an extremely strong local perturbation of the underlying collinear spin order, in particular along the **c** where the couplings are weak. Such perturbation can extend to several unit cells (Fig. 1B,D), become collective and give rise to a spiral if i) the Fe-Fe defects do not cluster within the **ab**-plane, i.e., if they are separated by a minimal average distance $\xi_{min}$, ii) the $J_{Cu-Fe} / J_{Fe-Fe}$ ratio is small enough and negative in order to guarantee strong local frustration, and iii) the concentration of Fe-Fe defects $n$ is low enough so as to avoid that impurity bonds dominate the order between successive planes ($n < J_{Cu-Fe} / J_{Fe-Fe}$)[32, 33].

The main prediction of this model is that *both*, the spiral ordering temperature $T_{spiral}$ and the magnetic modulation vector $q$ should increase linearly with $n$, as long as it remains at such low levels[32]. Interestingly, this is qualitatively verified in YBaCuFeO$_5$ samples prepared with different cooling rates, where the highest $q$ and $T_{spiral}$ values are observed for the samples quenched in liquid nitrogen (i.e., those with the largest Cu/Fe disorder and where $n$ is expected to be the highest, see Fig. 2A)[21]. It also suggests that the spiral ordering temperature could be further increased if values of $n$ slightly larger could be achieved. Since faster cooling rates are not easily achieved experimentally, we explore here an alternative strategy based in the combination of the highest possible Cu/Fe chemical disorder with a targeted lattice tuning of some magnetic exchanges.

**Adding the effect of the lattice**

Based in the previous discussion, the main candidates to influence the degree of magnetic frustration at constant Cu/Fe disorder are the NN couplings $J_{c1}$ and $J_{c2}$ along the *c*-axis, that we expect to display a negative correlation with the variation of the inter-bowtie separation ($d_1$) and the bowtie size ($d_1$), respectively (Fig. 1C)[21]. Since large changes in these two interatomic distances can be achieved through isovalent A-cation substitutions, we prepared two series of layered perovskites aimed to tune them separately. In the first one $Y^{3+}$ was replaced by trivalent 4f cations $Re^{3+}$ with growing ionic radius ($Re_{ionic}$) with the purpose of acting on $d_1$ (Re = Lu to Dy); in the second, $Ba^{2+}$ was partially substituted by different amounts of smaller $Sr^{2+}$ ($0 \leq x \leq 0.5$) with the aim of tuning $d_2$. In order to add-up the impact of both, the lattice and the Cu/Fe disorder, this last parameter was maximized in all the samples, which were quenched in liquid nitrogen after the last annealing (see Materials and methods section).

The impact of such $d_1$-tuning on $T_{spiral}$ is illustrated in Fig. 2B, which also shows the evolution of $T_{collinear}$ with the rare earth ionic radius in the $ReBaCuFeO_5$ series as determined from powder neutron diffraction (PND). The presence of the collinear and spiral phases in all samples was confirmed by the observation of magnetic Bragg reflections corresponding to $\mathbf{k_c}$ = (½ ½ ½) and $\mathbf{k_i}$ = (½ ½ ½±$q$) (Figs. 2D-E, figs. S1 and S2). The two transition temperatures change monotonically with $Re_{ionic}$ and display completely opposite behaviors, with $T_{collinear}$ decreasing and $T_{spiral}$ increasing for larger ionic radii. The lowest spiral ordering temperature corresponds to $LuBaCuFeO_5$ (213 K), and the highest to $DyBaCuFeO_5$ (312 K). This last value, the only one above RT (39 ºC), is 2 degrees higher that the $T_{spiral}$ reported in ref. 21 for quenched $YBaCuFeO_5$ (Fig. 2A).

The manipulation of $d_2$ has also a strong impact in both transition temperatures. This is illustrated in Fig. 2C, where the evolution of $T_{collinear}$ and $T_{spiral}$ in the $YBa_{1-x}Sr_xCuFeO_5$ family as a function of the Sr-content x is shown. PND measurements confirmed again the existence of collinear and spiral phases with propagation vectors $\mathbf{k_c}$ = (½ ½ ½) and $\mathbf{k_i}$ = (½ ½ ½±q) in all the materials investigated (Figs. 2F-G, figs. S1 and S2). We note that, contrarily to the predictions of ref. 26, $T_{collinear}$ decreases with x. $T_{spiral}$ displays the opposite behavior, reaching 375 K (102 ºC) for x = 0.4. Linear extrapolation indicates that both temperatures should merge at ~ 395 K for x ~ 0.49, giving rise to a paramagnetic-collinear-spiral triple point. For x = 0.5, i.e., just after this point, the sample enters into a different regime characterized by an abrupt change in the stability of the spiral state. As shown in Figs. 2F and S3, the magnetic reflections associated to the propagation vector $\mathbf{k_i}$ = (½ ½ ½±q) are still present, but they are extremely weak. Moreover, they coexist with a new, very strong reflection corresponding to the propagation vector $\mathbf{k_{c'}}$ = (½ ½ 0) and those of the $\mathbf{k_c}$ = (½ ½ ½) collinear phase. These observations suggest that the spiral state is abruptly suppressed and replaced by a new antiferromagnetic phase beyond the crossing of the $T_{collinear}$ and $T_{spiral}$ lines.

The temperature dependence of q for the ReBaCuFeO$_5$ and YBa$_{1-x}$Sr$_x$CuFeO$_5$ series together with that of YBaCuFeO$_5$ prepared using different cooling rates[21] are shown in Figs. 3A-C. For the samples with the lowest $T_{spiral}$ we observe a smooth, continuous decrease of q characteristic of second-order phase transitions. Such behavior is progressively replaced by an abrupt collapse of q in the samples with higher $T_{spiral}$, suggesting that the collinear-to-spiral transition changes from second- to first-order by approaching the $T_{spiral}$ = $T_{collinear}$ point (fig. S3). The evolution of the spiral plane inclination ($\varphi_G$) and the q value ($q_G$) at 10 K is shown in Figs. 3D-F. For the three series of samples we observe a clear, positive correlation among $T_{spiral}$, $\varphi_G$ and $q_G$. In each family the highest $T_{spiral}$ is observed for the sample with the largest $\varphi_G$, which reaches values close to 90º for Re = Dy. This corresponds to a perfect cycloid, indicating that the samples with the highest spiral ordering temperature have the potential of displaying the largest saturation polarization[15, 16]. Establishing if this is the case is out of the scope of this study and will require further experimental and theoretical work.

An interesting observation, common to all the samples investigated in this study, is the existence of a linear, universal relationship between the ordering temperature and the ground state periodicity of the spiral. This can be better appreciated in Fig. 4, where the evolution of $T_{spiral}$ with $q_G$ for the three series of samples is shown. We note that the extrapolation of the linear law towards low-$T_{spiral}$ values crosses the origin (i.e. $T_{collinear}$ = 0 for $q_G$ = 0), whereas in the high-$q_G$ side, $T_{collinear}$ and $T_{spiral}$ converge to a common value of ~ 395 K for $q_G$ ~ 0.18. The sudden disappearance of the spiral and the observation of a new magnetic phase close to this point suggest the existence of a limiting value for *both,* the spiral's periodicity and ordering temperature. In other words, 395 K is most probably the *highest* $T_{spiral}$ value that can be reached in Cu-Fe-based layered perovskites. To our best knowledge, it is the highest spiral ordering temperature reported for a transition metal oxide[34]. Moreover, it is comfortably far from room temperature, an important point regarding applications, and almost 100 K higher than the highest $T_{spiral}$ value previously reported for $YBaCuFeO_5$ using chemical disorder alone (310 K).

**Chemical disorder versus lattice tuning**

After demonstrating the success of our combined approach, we investigate in detail the associated changes in the crystal structure with the aim to getting additional insight on their link with the degree of magnetic frustration. We employ high resolution PND (fig. S6) to obtain precise values of the different structural parameters for the ReBaCuFeO$_5$ and YBa$_{1-x}$Sr$_x$CuFeO$_5$ series, that we compare with those obtained for YBaCuFeO$_5$ prepared using different cooling rates[21]. We check first the average Cu/Fe occupancies $n_{Cu}$ and $n_{Fe}$ of the bipyramidal sites, that we expected to be very similar owing to the quenching procedure used during the synthesis. The obtained values, shown in Figs. 5A-C and tables S3-6, confirm a high degree of disorder in all samples, which display average Cu/Fe occupancies very close to 50%.

The evolution of the lattice parameters and the tetragonal distortion $c/2a$ for the three series are summarized in Figs. 5D-L. Based in the results obtained for YBaCuFeO$_5$ prepared with different degrees of Cu/Fe disorder, we established in ref. 21 an empirical, positive correlation between $c/2a$ and $T_{spiral}$, and suggested the use of this parameter for $T_{spiral}$-control purposes. Although a similar correlation wss observed for the ReBaCuFeO$_5$ series (Fig. 5K), Fig. 5L reveals the opposite behavior in the Sr-substituted family. This suggests that the use of $c/2a$ for the control of magnetic frustration, of potential interest in thin film growth, cannot be generalized to all Cu/Fe layered perovskites.

The modification of the inter- and intra-bowtie distances $d_1$ and $d_2$ along the **c**-axis and their ratio $d_1/d_2$ for the three series are summarized in Figs. 5M-U. The longest $d_1$ and shortest $d_2$ values are systematically observed in the three families for the samples with the highest $T_{spiral}$ and lowest $T_{collinear}$ values. Moreover, the $d_1/d_2$ ratio displays a positive correlation with the spiral ordering temperature in the three series. This common behavior indicates that, contrarily to $c/2a$, the $d_1/d_2$ ratio is a good parameter for the control of magnetic frustration in Cu/Fe-based layered perovskites.

In spite of a common positive correlation between $T_{spiral}$ and $d_1/d_2$ ratio, Fig. 5 unravels the existence of large differences among the structural modifications in the three series. The smallest changes in the lattice parameters and the interatomic distances correspond to the YBaCuFeO$_5$ samples with different degree of Cu/Fe disorder. Interestingly, such tiny variations give raise to changes in $T_{spiral}$ of more than 150 K, much larger than in the ReBaCuFeO$_5$ (~100 K) and YBa$_{1-x}$Sr$_x$CuFeO$_5$ (~90 K) series where the variation of the lattice parameters and interatomic distances is much more pronounced. This is better appreciated in Fig. 5V, where the evolution of $T_{spiral}$ and with $d_1/d_2$ is displayed. This parameter, which reflects the relative strength $J_{c2}/J_{c1}$ of the couplings along the *c*-axis, changes very little in the YBaCuFeO$_5$ samples with different degree of disorder (0.41%). However, it increases by more than 1.7% between the two extremes of the YBa$_{1-x}$Sr$_x$CuFeO$_5$ family, and by 4.6% in the ReBaCuFeO$_5$ series. These results confirm the exceptional efficiency of chemical disorder alone for $T_{spiral}$ tuning. On the other side, our data show also that a comparable efficiency in terms of $T_{spiral}$ increase can be reached at fixed degree of chemical disorder by acting *exclusively* on the lattice. A particularly interesting observation is that the impact of both mechanisms on the spiral ordering temperature is additive: once the highest $T_{spiral}$ value is reached by manipulating the Cu/Fe disorder, it can be further increased through the maximization of the $d_1/d_2$ ratio. This allows $T_{spiral}$-tuning over unprecedentedly large temperature ranges, providing at the same time versatile tool for stabilizing spin spirals at temperatures well beyond RT.

**Experiment versus theory**

Having presented the experimental facts, we compare now our observations with the predictions of the random frustrating bonds (RFB) model proposed by Scaramucci and co-workers[32, 33]. The main prediction of this model is that both, $T_{spiral}$ and $q$ should increase with the concentration of Fe-Fe frustrating bonds $n$. Since $n$ is in turn correlated with the degree of average Cu/Fe disorder, such prediction explains well the behavior reported for the YBaCuFeO$_5$ samples prepared with different cooling rates[21]. However, it is difficult to conciliate with the huge $T_{spiral}$ changes observed in the YBa$_{1-x}$Sr$_x$CuFeO$_5$ and ReBaCuFeO$_5$ families, where the Cu/Fe disorder was kept approximately constant. A similar situation is observed with a second prediction, namely, the existence of a proportionality law between $T_{spiral}$ and $q_G$. Such linear law should be verified by samples with different degrees of disorder, and this is indeed the case for the YBaCuFeO$_5$ series prepared with different cooling rates[21] (blue full circles in Fig. 4). More intriguing is the fact that the very same law is also verified by the YBa$_{1-x}$Sr$_x$CuFeO$_5$ (green full pentagons) and ReBaCuFeO$_5$ families (red full squares), and even by the YBaCuFeO$_5$ single crystal from ref. 29 (black star) to a very good approximation. This unexpected observation cannot be explained within the framework of the RFB model, and calls for the explicit consideration of lattice's impact in the different magnetic exchanges.

A third point needing further theoretical work is the origin of the intersection of $T_{spiral}$ and $T_{collinear}$ for $q_G \sim 0.18$, and the sudden replacement of the spiral state by a second commensurate antiferromagnetic phase beyond this point. Given that it coincides with the smallest $d_2$ (i.e., largest $J_{c2}$) values of our study, the disappearance of the spiral could be tentatively ascribed to the point where the AFM couplings between Fe-Fe impurities overcome the FM coupling of the bipyramids occupied by Cu-Fe pairs. The result would be a magnetic structure were all NN couplings are AFM, in agreement with the observation of the new magnetic propagation vector $\mathbf{k_{c'}}$ = (½ ½ 0). However, it is not clear why this point should coincide with the intersection of the $T_{collinear}$ and $T_{spiral}$ lines. Another intriguing observation is the continuous evolution of the spiral-to-collinear phase transition from 2$^{nd}$ to 1$^{st}$ ordering by approaching the paramagnetic-collinear-spiral triple point. Information about the behavior of the different magnetic phases beyond the crossing of the $T_{collinear}$ and $T_{spiral}$ lines should help to understand its exact nature, and in particular, to clarify whether we could be in presence of a Lifschitz point[35, 36]. We note also that its increased entropy at the crossing makes this point potentially interesting for the search of exotic magnetic states in presence of external perturbations such as magnetic fields or elevated pressures[35, 37, 38].

## Towards spiral design in layered perovskites

An important outcome of the structural property correlations presented in previous sections is that they enable to propose a simple set of rules for the design of high-temperature spin-spirals in other AA'BB'O$_5$ layered perovskites with B-site disorder. Besides maintaining electric neutrality, the most critical point is to identify B/B' cation pairs of comparable sizes, affinity for square-pyramidal coordination, able to produce weak, alternating FM and AFM couplings along *c*-axis, and with a *very* large value for *one* of the two $J_{BB}$ or $J_{B'B'}$ couplings. The choice is mostly limited by the relatively small number of pairs resulting in FM coupling within the pyramids, but among 3d transition metals, $Cu^{2+}/Cu^{3+}$, $Cu^{2+}/Co^{3+}$(high-spin), $Cu^{2+}/Mn^{3+}$ or $Mn^{2+}/Co^{3+}$(low-spin) could be possible candidates if $J_{BB}$ and $J_{B'B'}$ have the appropriate absolute values. DFT estimations of the magnetic exchanges, in particular for 4d and 5d pairs, will be highly desirable, and shall increase the number of potential candidates. The choice of the A/A'-cations requires ionic radii different enough to stabilize the layered perovskite structure. However, the highest $T_{spiral}$ values can be achieved when the ionic radius of the A-cation close to the apical oxygen is small, and the $r^A_{ionic} - r^{A'}_{ionic}$ difference as small as possible (Fig. 5). Concerning thin film design, we saw that $c/2a$ alone is not a good parameter for $T_{spiral}$ control. Hence, high $T_{spiral}$ values can be a priori stabilized under both, tensile or compressive strain as long as the rules concerning the choice of the A and B cations are fulfilled.

We can also speculate on the expected behavior of $T_{spiral}$ under external pressure. According to our findings, this will depend on the behavior of $d_1/d_2$, i.e., on the relative compressibility of these two distances. In general, bonds involving cations with high coordination or low valence expand or compress faster than those with cations in low coordination or with high valence[39, 40]. $d_2$, involving 12 ($Ba^{2+}$-$O^{2-}$) bonds, is thus expected to compress faster than $d_1$, where only 8 ($R^{3+}$-$O^{2-}$) bonds contribute. Since this will result in a smaller $d_1/d_2$ ratio, a decrease of $T_{spiral}$ is thus predicted under the application of external pressure.

To summarize, we have experimentally validated an emerging route for stabilizing spin spirals up to temperatures far beyond RT. Our approach, based in the combination of chemical disorder with a targeted lattice tuning of some magnetic exchanges, takes full advantage of the additivity of these two mechanisms. As a result, the stability domain of the spiral magnetic order can be tuned over unprecedentedly large temperature ranges.

In the particular case of Cu/Fe-based layered perovskites we demonstrate that $T_{spiral}$ values close to 400 K can be reached using this strategy. This value is ~100 K higher than using chemical disorder alone[21], and comfortably far beyond RT. We also reveal the existence of a linear, universal relationship between the spiral's ordering temperature and periodicity, a paramagnetic–collinear-spiral triple point, as well as several correlations between the spiral properties and some structural parameters. We discuss these results on the light of a recently proposed random frustrating bonds based mechanism, and propose a simple set of rules for the design of magnetic spirals in isostructural layered perovskites by using external pressure, chemical substitutions or epitaxial strain. Besides overcoming one of the main stoppers for applications, these results could accelerate the discovery of other materials featuring spiral phases stable well beyond RT, paving the way towards the long-sought technological use of magnetic spirals in spintronics devices.

## Materials and Methods

**Materials synthesis**

The $ReBaCuFeO_5$ and $YBa_{1-x}Sr_xCuFeO_5$ polycrystalline samples were prepared using the solid state reaction method. High-purity $Re_2O_3$ or $Y_2O_3$ (99.995 - 99.998%, Aldrich / Alfa Aesar), $SrCO_3$ (99.995%, Aldrich), $BaCO_3$ (99.997%, Alfa Aesar), CuO (99.995%, Alfa Aesar) and $Fe_2O_3$ (99.998%, Alfa Aesar) powders were used as starting materials. Dehydrated rare-earth oxide powders were obtained by heating them at 1223 K for 15 hours. The required stoichiometric amounts of $Re_2O_3$, $SrCO_3$, $BaCO_3$, CuO and $Fe_2O_3$ were then weighted, thoroughly ground and heated at a rate of 300 K/h up to the synthesis temperature $T_s$ (= 1323 – 1430 K), that was optimized for every sample using thermogravimetric analysis under He/O$_2$ gas flow. The powders were then annealed during 50 hours at this temperature under oxygen gas flow. The obtained materials were cooled in the furnace down to RT, thoroughly grounded again, pressed into pellets, and sintered at $T_s$ for another 50 hours. After this treatment, all samples were quenched into liquid nitrogen. Small pieces were kept solid for their use in magnetization measurements, and the rest was pulverized and subsequently employed in PND experiments. The phase purity of all samples was checked using laboratory powder x-ray diffraction (XRD) at RT using a Brucker D8 Advance diffractometer with Cu K$_\alpha$ radiation. All samples were very well crystalized and free of impurities within the limit of this technique (~1%).

**Magnetic susceptibility**

DC magnetization (M) measurements were performed in a superconducting quantum interference device magnetometer (MPMS-XL 7T, Quantum Design). Small pellets ($D \sim$ 3 mm, $H \sim$ 1 mm, $m \sim$ 15-25 mg) from the same batches as the samples used for the PND measurements were measured between 2 and 400 K by heating under a magnetic field of $\mu_0 H = 0.5$ T after being cooled down to 2 K in zero field. The magnetization was further measured between 300 and 600 K using a high temperature insert. The samples, mounted on transparent drinking straws for the measurements below 400 K, were dismounted and wrapped in aluminum foil as described in ref. 41 for the measurements above this temperature. The signal from the aluminum foil, measured separately, was found to be temperature-independent and negligible when compared with the sample's magnetization. For the samples with magnetic Re$^{3+}$ cations, the magnetic transitions from Cu/Fe sublattice are difficult to observe in the magnetization data due the large paramagnetic contribution of the Re$^{3+}$ moments, and they have been missed in some previous studies[42]. However, they can still be tracked in the first derivatives of the DC inverse susceptibility $1/\chi^{DC}$ = H/M (fig. S1). The values of $T_{collinear}$ and $T_{spiral}$ derived from this technique correspond to the midpoint of the step-like anomalies in the $1/\chi^{DC}$ derivative, which coincides with the $\chi^{DC}$ maximum in YBa$_{1-x}$Sr$_x$CuFeO$_5$ series.

It is worth mentioning that, although the ReBaCuFeO$_5$ samples investigated in this work were also synthesized in the past[43, 44], the existence of magnetic transitions has only been reported for a few of them[42, 45, 46, 47, 48], and there is no apparent systematics between the reported ordering temperatures and the Re ionic radii. A similar situation was observed in the YBa$_{1-x}$Sr$_x$CuFeO$_5$ series, where the presence of spiral phases was either disproven[49], or not demonstrated[50] in previous works. As shown in Fig. 2C, this is not the case in this study, where the collinear and spiral phases are clearly observed for *all* samples. Moreover, $T_{collinear}$ and $T_{spiral}$, change in a monotonic way as a function of both, the Re ionic radius and the Sr content x.

**Differential scanning calorimetry**

The paramagnetic to collinear phase transition at $T_{collinear}$ was also measured by differential scanning calorimetry (DSC) for ReBaCuFeO$_5$ and YBa$_{1-x}$Sr$_x$CuFeO$_5$ powder samples by using a NETZSCH DSC 204F1 device. Fine powders (~ 30 mg mass) were sealed into aluminum crucibles and heated up to 670 K with a rate of 15 K/min under Ar gas flow. The data were acquired during the heating process (see fig. S2).

**Powder neutron diffraction**

Powder neutron diffraction measurements were carried out at the Swiss Neutron Source SINQ of the Paul Scherrer Institute in Villigen, Switzerland, and the Institut Laue-Langevin in Grenoble, France. The ReBaCuFeO$_5$ and YBa$_{1-x}$Sr$_x$CuFeO$_5$ powder samples were introduced in cylindrical vanadium cans with (D = 6 mm, H = 5 cm) and mounted on the stick of a helium cryostat (2 - 300 K) and a cryofurnace (2 - 500 K). PND patterns were continuously recorded at the powder diffractometer DMC (Pyrolitic Graphite (002), $\lambda$ = 2.458 Å and $\lambda$ = 4.507 Å, $2\theta_{step}$ = 0.1°) while ramping the temperature from 10 to 480 K. Longer acquisitions for magnetic structure refinements were made at 10 K. High resolution patterns at both 10 K and 300 K were also recorded at the powder diffractometer HRPT (Ge (822), $2\theta_{step}$ = 0.05°, $\lambda$ = 1.154 Å) and D2B (Ge (335), $2\theta_{step}$ = 0.05°, $\lambda$ = 1.594 Å). The wavelengths and zero offsets were determined using a NAC reference powder sample. In the case of DyBaCuFeO$_5$ the impact of the relatively large neutron absorption cross section of natural dysprosium (994(13) barn) in the data was corrected by including in the refinements the µR values determined experimentally for the different wavelengths. The background from the sample environment was minimized through 5' collimators (D2B), and an oscillating radial collimator (HRPT). The values of $T_{collinear}$ and $T_{spiral}$ derived from this technique correspond respectively, to the set-up and the maximum of the magnetic Bragg reflection (½ ½ ½), that roughly coincide with the values determined from $\chi^{DC}$ and DSC (see figs. S1 and S2, tables S1 and S2).

**Data analysis**

All diffraction data were analyzed using the Rietveld package FullProf Suite[51]. The structural refinements for ReBaCuFeO$_5$ were carried out by combing both HRPT and D2B data sets recorded at room temperature and 10 K. For YBa$_{1-x}$Sr$_x$CuFeO$_5$, HRPT data recorded at room temperatures and 10 K were used for the fits of the crystal structure (fig. S6). The magnetic refinements were performed on DMC data recorded at 10 K by fixing the structural parameters as determined from refinements of HRPT and D2B patterns (fig. S4). We used the non-centrosymmetric space group *P4mm* for the description of the crystal structure. The choice was motivated by the fact that, contrarily to the centrosymmetric space group *P4/mmm*, *P4mm* enables to refine the occupation of the split Cu and Fe sites.

The Cu/Fe disorder was described by splitting the atomic position (½ ½ $z$) inside of the two pyramidal sites of the unit cell as shown in the tables S3 to S6. In the fully disordered case both positions are equally occupied in all pyramids (i.e., $n_{Fe} = n_{Cu} = 50\%$), whereas full Cu/Fe order corresponds to $n_{Cu} = 100\%$, $n_{Fe} = 0$ (or $n_{Cu} = 0$, $n_{Fe} = 100\%$). As shown in tables S3 to S6, the samples investigated in this work are very close to the first scenario, i.e., of $|n_{Cu} - n_{Fe}|$ very close to zero. Note that Rietveld fits are not sensible to occupational correlations. The Cu/Fe disorder is thus assumed to be random, and the refined $n_{Cu}$ and $n_{Fe}$ *values* represent the average values over the full sample.

Anisotropic Debye-Waller factors were used for all atoms with exception of Y/Re (nearly isotropic, see ref. 21), Cu and Fe. The $z$ coordinates of the two basal oxygen sites O2 and O2' were refined separately but their mean-square displacements (MSD) were restricted to have the same value (tables S4 to S6). No signature of interstitial oxygen at the (½, ½, $z$) atomic position could be observed, indicating that deviations from the sample's formula stoichiometry, if any, were within the detection limit of PND in all the samples investigated. For the YBa$_{1-x}$Sr$_x$CuFeO$_5$ sample with x = 0.5 we could not find any evidence of alternating Ba/Sr layers, recently predicted for $x \geq 0.5^{26}$ and easy to identify due to the associate doubling of the *c* lattice parameter.

The collinear and spiral magnetic structures were described according to the models reported in ref. 18. The best fits obtained at 10 K for all samples are shown in fig. S4. The ratio between the $Fe^{3+}$ ($3d^5$ HS) and $Cu^{2+}$ ($3d^9$) magnetic moments was restricted to be the same than their free ion, spin-only values (5:1). In the temperature regions where the collinear and spiral phases coexist, the Fe and Cu magnetic moments were restricted to have the same value and the same inclination with respect to the *ab*-plane in the two magnetic phases.

**Acknowledgements:** We thank A. Scaramucci, M. Müller, Ch. Mudry, N. Spaldin, M. Shi, Z. Wang, M. Radovic, J. Mesot, R. Sibille and T. Mazet for fruitful discussions. We acknowledge the allocation of beam time at the Swiss Neutron Source SINQ (HRPT and DMC diffractometers) and Institut Laue-Langevin in Grenoble, France (D2B diffractometer). E. C. acknowledges support from the Danish Research Council for Science and Nature through DANSCATT. **Funding:** This work was supported by the Swiss National Science Foundation (Grants No. 200021_141334 and 206021_139082). **Author contributions:** T.S. and M.Me. conceived and led the project, T.S, and M.Mo. synthesized the samples, T.S., E.C., M.Me., D.S., and M.T.F.-.D. performed the PND measurements, T.S. measured the magnetic susceptibility and DSC, T.S., and M.Me. analyzed all the experimental data. T.S. and M.Me. wrote the paper with the input from all authors. **Competing interests:** The authors declare no competing financial interests. **Data and materials availability:** Raw powder diffraction data were generated at the SINQ (Paul Scherrer Institut, Switzerland) and ILL (Grenoble, France). Derived data supporting the results of this study are available from the corresponding authors. **Code availability:** FullProf Suite is available free of charge at https://www.ill.eu/sites/fullprof

**Figures**

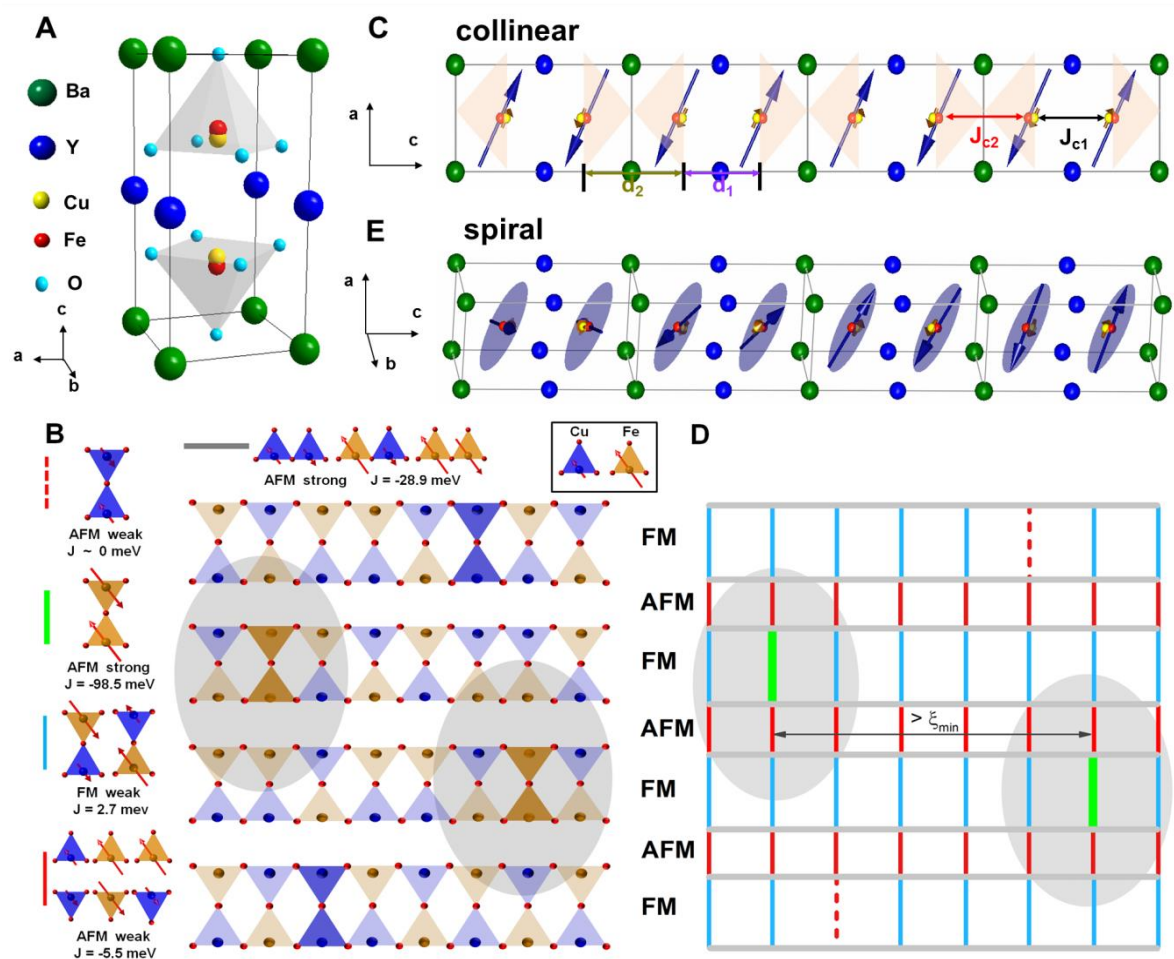

**Fig. 1. Link between Cu/Fe chemical disorder and magnetic order in YBaCuFeO$_5$.** **(A)** Crystal structure of YBaCuFeO$_5$ with Cu/Fe disorder in the bipyramidal sites. **(B),** Schematic representation of the correlated Cu-Fe chemical disorder in YBaCuFeO$_5$, characterized by the presence of randomly-disordered Cu-Fe FM 'dimers'. **(C)** Magnetic structures of the commensurate collinear phase. $J_{c1}$ and $J_{c2}$ denote respectively the nearest neighbor AFM and FM couplings along the *c*-axis. **(D)** The representation of (B) in terms of weak, alternating AFM/FM bonds. The impact of a few, randomly distributed strong AFM Fe-Fe 'defects' is illustrated by the grey ellipses, which delimit the regions where the Cu-Fe spins loss their collinearity. The cross-talking of such regions eventually gives rise to a spiral if their separation is smaller than $\xi_{min}$. Note that the same number of weakly interacting Cu-Cu defects is necessary in order to preserve the material's stoichiometry. The *J*-values are those of ref. 32 **(E)** Magnetic structure of the incommensurate spiral phase.

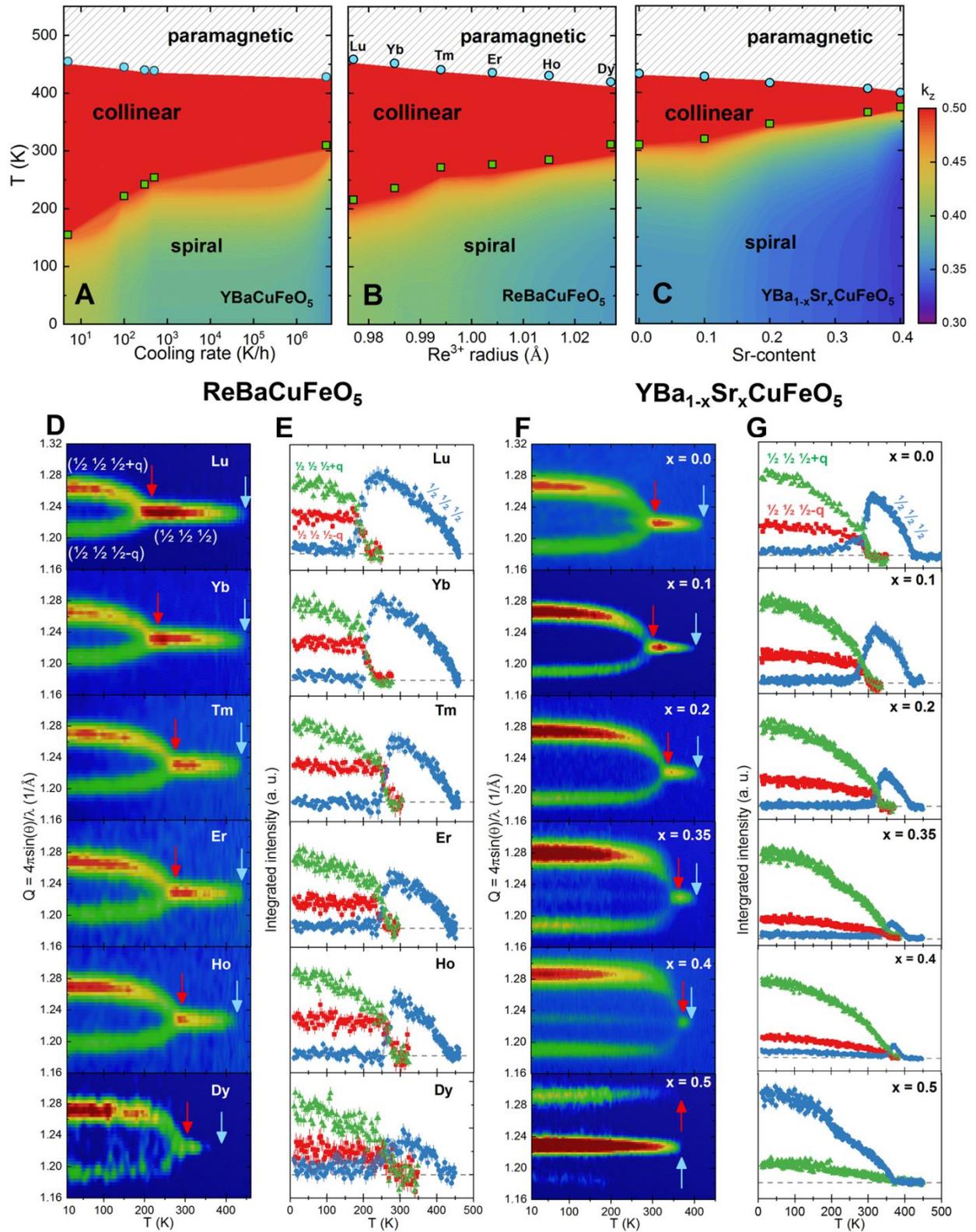

**Fig. 2 Stability of the collinear and spiral phases with different tuning parameters. (A-C)** Evolution of $T_{collinear}$ and $T_{spiral}$ using different tuning parameters: (A) YBaCuFeO$_5$ prepared with different cooling rates (adapted from ref. 21); (B) Replacement of Y$^{3+}$ by a rare-earth cation (Re$^{3+}$); (C) Replacement of Ba$^{2+}$ by Sr$^{2+}$. The $T_{collinear}$ and $T_{spiral}$ values are those determined from PND. The background color represents the z-component of magnetic propagation vector ($k_z$) which changes from ½ in the collinear phase to ½±q in the spiral phase (only ½-q is used in the figures). **(D-G)** Signature of the collinear and spiral phases in the PND patterns. The contour maps show the temperature dependence of the positions and the intensities of the commensurate (½ ½ ½) magnetic Bragg reflection and the incommensurate (½ ½ ½±q) satellites for the ReBaCuFeO$_5$ (D) and YBa$_{1-x}$Sr$_x$CuFeO$_5$ (F) series. The blue and red arrows indicate the ordering temperatures $T_{collinear}$ and $T_{spiral}$, respectively. The integrated intensities are shown in (E) for ReBaCuFeO$_5$, and (G) for YBa$_{1-x}$Sr$_x$CuFeO$_5$. For the Sr-doped sample with x = 0.5, the "minus" incommensurate satellites (½ ½ ½-q) was extremely weak (see also fig. S3). Hence, only the integrated intensity of the "plus" satellite (½ ½ ½+q) is shown. The error bars of the integrated intensities are the standard deviations obtained from 3-Gaussian least-square fits of the (½ ½ ½) Bragg reflection and its incommensurate satellites carried out using the Igor software.

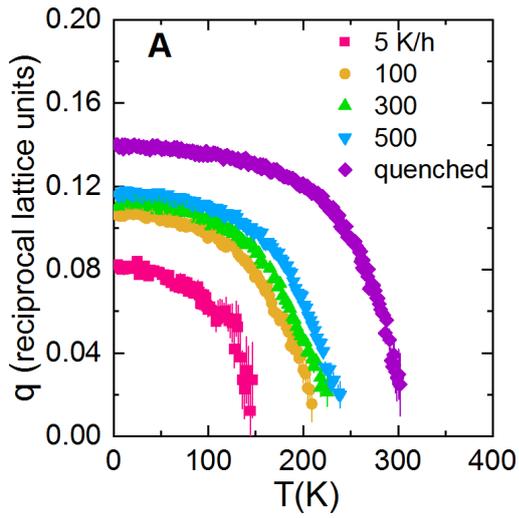
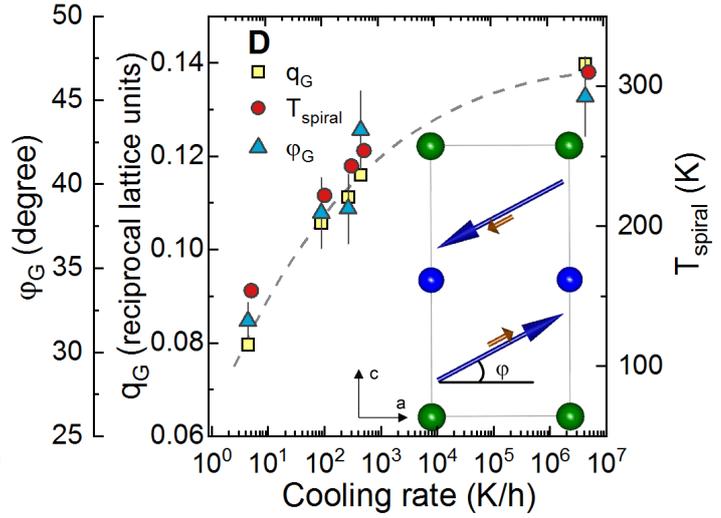
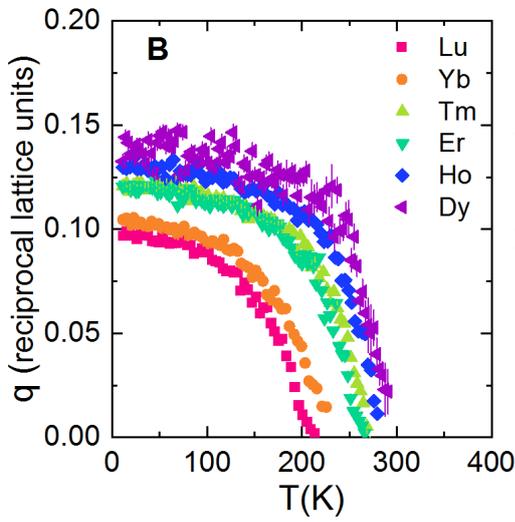
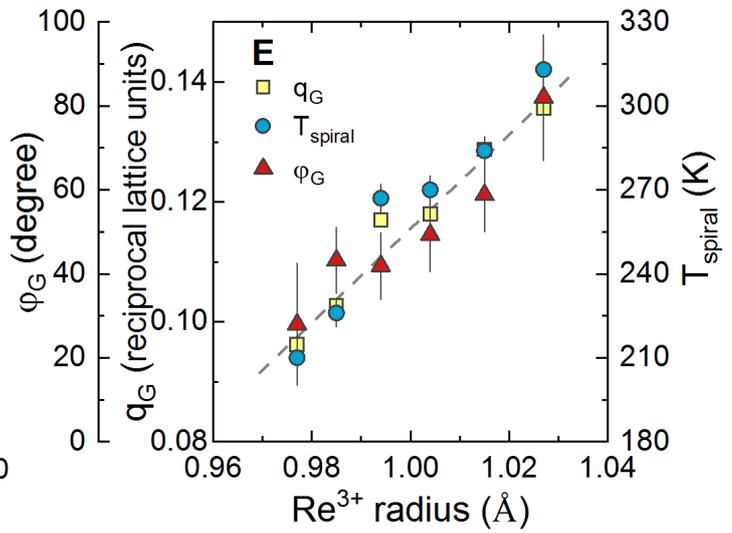
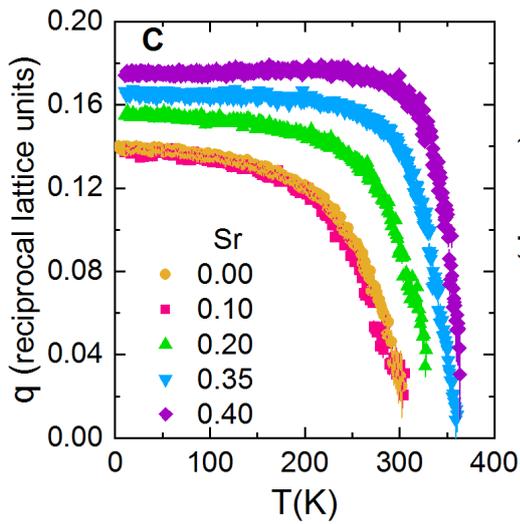
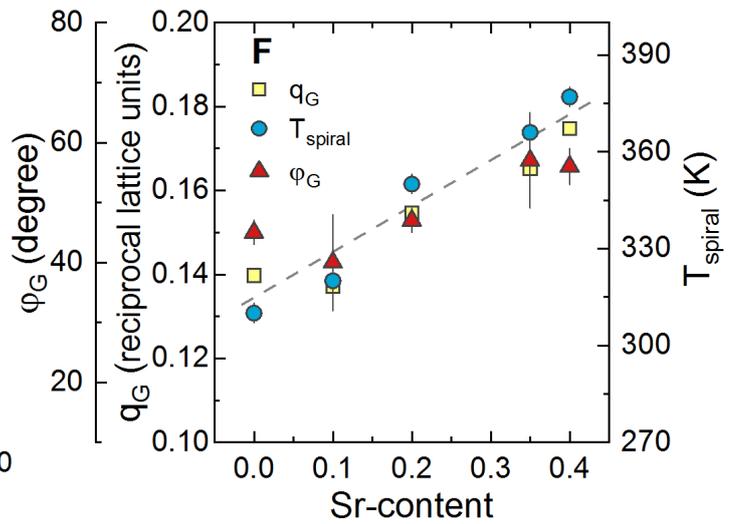

**Fig. 3. Evolution of magnetic spiral properties with different tuning parameters.**
**(A-C)** Temperature dependence of the magnetic modulation vector $q$ in YBaCuFeO$_5$ using different tuning parameters: (A) Cooling rate (adapted from ref. 21). (B) Replacement of Y$^{3+}$ by a rare-earth cation (Re$^{3+}$). (C) Replacement of Ba$^{2+}$ by Sr$^{2+}$. **(D-F)** Evolution of $T_{spiral}$ and the ground state values (10 K) of the magnetic modulation vector ($q_G$), and the inclination of the spiral rotation plane ($\varphi_G$) using different tuning parameters: (D) Cooling rate (adapted from ref. 21). (E) Replacement of Y$^{3+}$ by a rare-earth cation (Re$^{3+}$). (F) Replacement of Ba$^{2+}$ by Sr$^{2+}$. The inset in (D) shows the definition of the inclination angle $\varphi_G$. The dashed lines are guides for the eyes and indicate the positive correlation among $T_{spiral}$, $q_G$, and $\varphi_G$. The values of ReBaCuFeO$_5$ and YBa$_{1-x}$Sr$_x$CuFeO$_5$ families were extracted from Rietveld fits of the DMC data recorded at 10 K. The error bars of $q_G$ and $\varphi_G$, listed in tables S3 and S4, are the standard deviations obtained from the fits of the magnetic structure using the FullProf Suite Rietveld package[51].

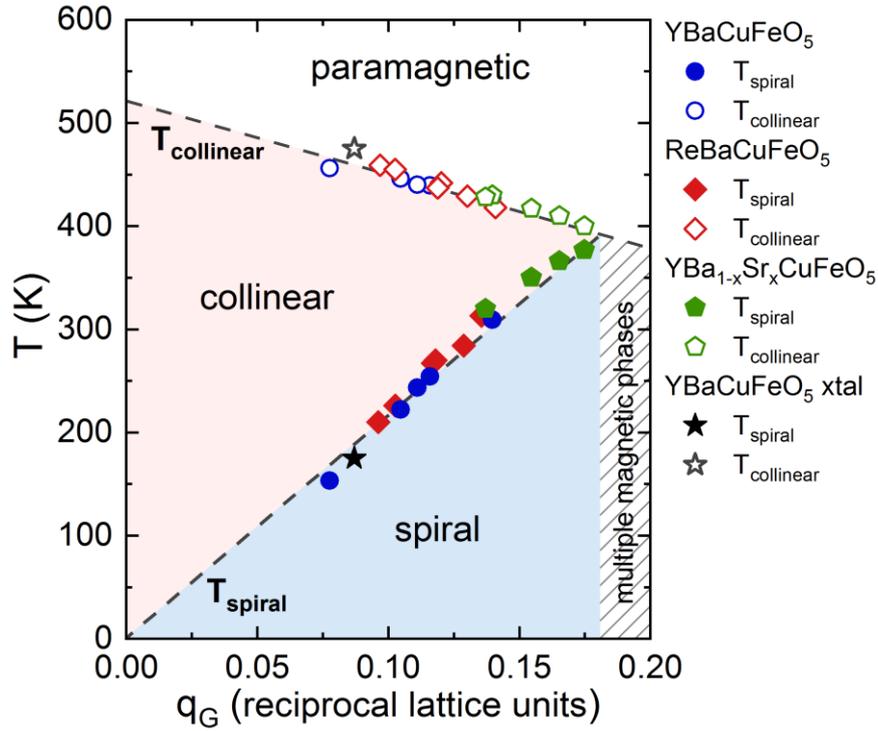

**Fig. 4. Universal relationship between $T_{spiral}$ and $q_G$ in AA'CuFeO$_5$ layered perovskites**. Correlation between $T_{spiral}$ and the ground state magnetic modulation vector $q_G$ in three families of AA'CuFeO$_5$ layered perovskites: YBaCuFeO$_5$ prepared with different cooling rates (full blue-symbols, data from ref. 21), ReBaCuFeO$_5$ with Re = Lu to Dy (full red-symbols), and YBa$_{1-x}$Sr$_x$CuFeO$_5$ with $0 \leq x \leq 0.5$ (full green-symbols). The full black star corresponds to the ($q_G$, $T_{spiral}$) point reported for a YbaCuFeO$_5$ single crystal in ref. 29. The $T_{collinear}$ values of all samples (open symbols) are also shown in order to illustrate its convergence with $T_{spiral}$ at $q_G \sim 0.18$. The $T_{spiral}$ and $T_{collinear}$ values are those determined from PND (tables S1 and S2). The error bars of $q_G$, listed in Supplementary tables S4 and S6, are the standard deviations obtained from the fits of the magnetic structure using the FullProf Suite Rietveld package[51], and they are smaller than the marker's size.

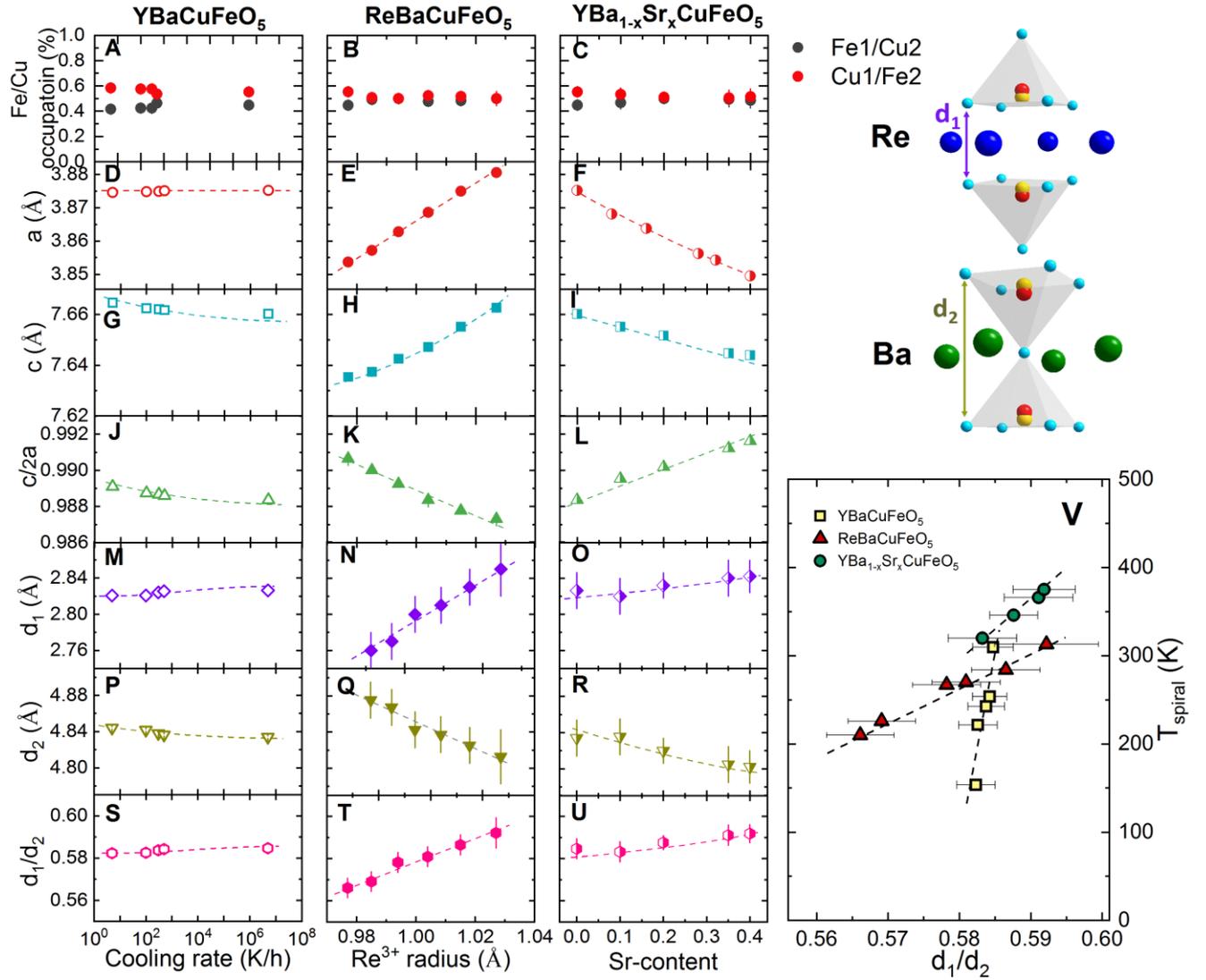

**Fig. 5**. **Link between crystal structure and magnetic ordering temperatures. (A-U)** Evolution of selected structural parameters as functions of the cooling rate (YBaCuFeO$_5$, data from ref. 21), the rare earth ionic radius (ReBaCuFeO$_5$), and the Sr content (x) (YBa$_{1-x}$Sr$_x$CuFeO$_5$). (A-C) Average Cu/Fe occupation in the pyramids. (D-I) Lattice parameters **a** and **c**. (J-L) Tetragonal distortion $c/2a$. (M-O) Distance $d_1$ between the bipyramidal layers along the **c**-axis. (P-R) Thickness $d_2$ of the bipyramidal layers. (S-U) $d_1/d_2$ ratio. **(V)** Evolution of $T_{spiral}$ with $d_1/d_2$ ratio. The yellow, red, and green colors correspond respectively to YBaCuFeO$_5$ prepared with different cooling rates, ReBaCuFeO$_5$ (Re = Lu - Dy), and YBa$_{1-x}$Sr$_x$CuFeO$_5$ (x = 0 - 0.4). The $T_{spiral}$ and $T_{collinear}$

values are those determined from PND (tables S1 and S2). The dashed lines are guides to the eyes. All structural parameters were extracted from the Rietveld fits of the high resolution PND data at RT (tables S3 and S5). The error bars are the standard deviations provided by the FullProf Suite Rietveld package[51].